\documentclass[11pt]{article}

\pdfoutput=1

\usepackage[T1]{fontenc}
\usepackage[latin9]{inputenc}
\usepackage[a4paper]{geometry}
\usepackage[active]{srcltx}
\usepackage{amsmath}
\usepackage{amssymb}
\usepackage{esint}
\usepackage{xcolor}
\usepackage{multicol}
\usepackage{graphicx}
\usepackage{dsfont}
\usepackage{mathtools}
\usepackage{caption}
\allowdisplaybreaks[1]

\usepackage[symbol]{footmisc}

\newcommand{\g}{\gamma}

\newcommand{\del}{\partial}
\newcommand{\be}{\begin{equation}}
\newcommand{\ee}{\end{equation}}
\newcommand{\bea}{\begin{eqnarray}}
\newcommand{\eea}{\end{eqnarray}}

\newcommand{\nn}{\nonumber}
\newcommand{\G}{\Gamma}

\makeatletter
\usepackage{textcomp}

\pdfoutput=1 % if your are submitting a pdflatex (i.e. if you have
\usepackage{jheppub}% for details on the use of the package, please
\usepackage{etoolbox}% http://ctan.org/pkg/etoolbox
    
    \patchcmd{\maketitle}{\@fpheader}{}{}{}

\usepackage{amsfonts}

\title{\boldmath Carroll fermions from null reduction: A case of good and bad fermions}

\author[]{Sucheta Majumdar${}^{1,2}$, Aditya Sharma${}^{3,4}$, Sourav Singha${}^5$}

\affiliation{${}^1$ Center for Space, Time and the Quantum, 13288 Marseille, France}
\affiliation{${}^2$Aix-Marseille Univ., Centre de Physique Th\`eorique, 13288 Marseille, France}
\affiliation{${}^3$Facultad de Ingenier\'{i}a, Universidad San Sebast\'{i}an, sede Valdivia, General Lagos 1163, Valdivia 5110693, Chile}
\affiliation{${}^4$Centro de Estudios Cient\'{i}ficos (CECs), Arturo Prat 514, Valdivia, Chile}
\affiliation{${}^5$Department of Physics, Indian Institute of Technology Kharagpur, Kharagpur 721302, India}
\emailAdd{sucheta.majumdar@cpt.univ-mrs.fr\\
ext.aditya.sharma@uss.cl , adityasharma.theory@gmail.com\\
singhasourav000@gmail.com}
\preprint{}

\abstract{
We derive Carrollian fermionic actions using the null reduction method from Bargmann spacetimes. In the Lorentzian light-cone formulation, the Dirac spinor naturally decomposes into dynamical and constrained degrees of freedom $-$ the so-called `good' and `bad' fermions $\Psi_{(\pm)}$. These light-cone projections are intrinsically adapted to the null frame and, unlike the chiral decomposition into left- and right-handed spinors $\Psi_{L(R)}$, are valid in arbitrary spacetime dimensions, both even and odd. As in the case of bosons, the magnetic Carroll sector for fermions is governed by the dynamical modes of the parent theory, while the electric sector arises from the constrained modes. Upon deforming to a Bargmann spacetime, these constraints are removed, promoting the `bad' fermions to dynamical modes that describe the electric Carroll fermions. We construct the Clifford algebra on the Carroll manifold through its embedding in the ambient Bargmann manifold, and obtain both electric and magnetic Carroll fermion actions from a \textit{single} Bargmann-invariant Dirac action. We analyze the canonical structure of both theories, establish their invariance under Carroll transformations, and compute the corresponding two-point functions, which exhibit the expected behavior in both sectors. We conclude with some comments on the quantization of these Carrollian theories.}

\makeatother

\begin{document}
\maketitle \flushbottom

\section{Introduction}
The past decade has seen a significant upsurge in understanding various aspects of the limit where the speed of light, $c\to 0$ (see e.g~\cite{Cardona:2016ytk, Bagchi:2019clu, Donnay:2019jiz, Banerjee:2020qjj, Henneaux:2021yzg, Perez:2021abf,Baiguera:2022lsw,Mehra:2023rmm,Sharma:2025rug} and references therein).  Geometrically, the vanishing of $c$ corresponds to the collapse of the light cone onto the temporal axis and renders the space absolute under boosts transformations. This alters our understanding of causality and a stricter notion of ultra-locality emerges, where the two events become causal only if they happen at the same spacetime point. While $c\to 0$ was originally dubbed as ultra-local limit~\cite{Klauder:1971zz, Klauder:2000ud, SenGupta:1966qer}, the pseudonym `Carroll' has resonated with the theorists~\cite{Levy-Leblond:1965dsc,Bacry:1968zf}. From a field-theoretic perspective, the implications of this ultra-locality are profound and it dictates the structure of correlation functions, which typically collapse into delta-function distributions between spatially separated points~\cite{Chen:2021xkw, Bagchi:2022emh, Donnay:2022wvx}.
\par
Arguably, the most prominent motivation for the recent flurry of research into Carrollian field theories is the idea of Carrollian holography which asserts that a Carrollian conformal field theory may serve as a holographic dual to the gravity in asymptotically flat spacetimes~\cite{Alday:2024yyj, Mason:2023mti, Donnay:2022aba,Hartong:2025jpp}. These dual field theories also go by the name of BMS-invariant field theories as it has been know for quite a while now that the BMS group is isomorphic to the conformal Carrollian group~\cite{Duval:2014uva}. Apart from the holographic motivation, Carrollian symmetry has also been realized in other physical settings such as hydrodynamics, black hole physics, string theory and many others e.g.~\cite{Duval:2017els,Ciambelli:2018wre,Freidel:2022bai,Gray:2022svz,Marsot:2022imf, Perez:2023uwt, Athanasiou:2024lzr,Kolekar:2024,Figueroa-OFarrill:2025njv}. A more comprehensive review on this subject can be found in~\cite{Bergshoeff:2022eog, Bagchi:2025vri,Ciambelli:2025unn,Nguyen:2025zhg,Ruzziconi:2026bix} and references therein.
\par
The interest in ultra-local field theories goes back to the work of Klauder in 1970's~\cite{Klauder:1971zz,Klauder:2000ud}, where scalar field theory models devoid of spatial gradients were studied; these were only later realized to be the electric sector of Carrollian field theories. However, in recent times, a significant effort has been made to construct various field theory models consistent with Carrollian symmetries in general, if not conformal~\cite{Sharma:2025rug, Mehra:2023rmm, Ekiz:2025hdn, Banerjee:2023jpi}. An interesting aspect of Carrollian field theories is that they manifest in two forms, namely electric and magnetic sector. This nomenclature is based on how the field(s) under consideration are scaled, akin to the construction of Galilean field theories (e.g~\cite{LeBellac:1973unm, Duval:2014uva, Banerjee:2022uqj, Santos:2004pq, Sharma:2023chs, Banerjee:2022eaj}). However, beyond scaling arguments, the `electric' and `magnetic' labels can also be ascribed a geometric basis dictated by how a Lorentzian theory is null-reduced~\cite{Duval:2014uoa}.  While the method of null reduction is well-established in the literature and has been extensively employed to construct Galilean-invariant field theories, its Carrollian counterpart has been developed quite recently in~\cite{Chen:2023pqf} and~\cite{Majumdar:2025ju}. The two approach however, differ from each other. While~\cite{Chen:2023pqf} describes how the electric and magnetic sectors of Carrollian theory originate from a different relativistic actions in one higher dimension, the approach in~\cite{Majumdar:2025ju} deforms the relativistic theory in one higher dimension to construct the two sectors from a single relativistic Lagrangian. Over the years, various other techniques have also been developed to construct field theory models consistent with Carrollian symmetries such as expansion method where the field content is expanded in the powers of $c$~\cite{deBoer:2023fnj,deBoer:2021jej,Bagchi:2019clu}, seed Lagrangian method~\cite{Bergshoeff:2022qkx} and bootstrapping techniques~\cite{Afshar:2024llh,Marotta:2025qjh}.
\par
In this paper, we construct field theory models for Carrollian fermion using~\cite{Majumdar:2025ju}. The subject of Carrollian fermions is  particularly interesting, as existing models in the literature either predominantly stem from pure symmetry arguments~\cite{Bagchi:2022eui, Mele:2023lhp, Ara:2024fbr,Grumiller:2025rtm, Ekiz:2025hdn} or, even in cases where a limiting procedure is employed, the parent Lagrangian is typically not the standard Dirac Lagrangian~\cite{Bergshoeff:2023vfd}. Additionally, these models employ chiral projections restricting the spinor properties to depend on the dimensionality of the spacetime.  The question of whether field theory models for Carrollian fermions can be constructed directly from the standard Dirac Lagrangian arises naturally for two principal reasons. First, by analogy with the bosonic theories, existing Carrollian scalar and gauge field models are known to descend systematically from parent Lorentzian Lagrangians. Second, there is a compelling geometric motivation: the null reduction of a Bargmann manifold along one of its null directions inherently induces a Carrollian structure~\cite{Duval:2014uoa}. Thus, despite significant work, a first-principles construction of Carrollian fermions starting from Dirac Lagrangian has remained elusive. 
\par
\begin{figure}[h]
    \centering
    \includegraphics[width=0.9\textwidth]{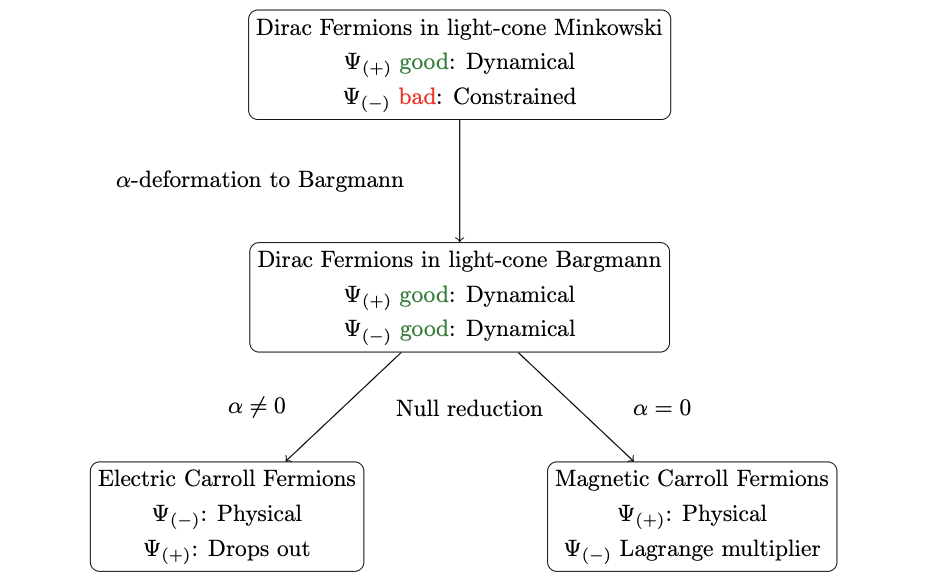}
    \caption{Carroll fermions from null reduction of light-cone Dirac fermions}
    \label{fig:Schematic}
\end{figure}

We address this issue in this paper by employing the method of null reduction.  By taking the conventional Bargmann Dirac Lagrangian as our starting point, we null-reduce the theory to construct both the electric and magnetic sectors of Carrollian fermions. Our construction rely on reformulating the Dirac Lagrangian in Lorentzian light-cone variables and subsequently decomposing the spinor into its light-cone projections, $\Psi_{(\pm)}$; the so-called `good' and `bad' fermions~\cite{Kogut:1969xa,Bjorken:1970ah,Chang:1972xt}. This framework naturally isolates the dynamical degrees of freedom from the constrained ones.     Analogous to the bosonic case~\cite{Majumdar:2025ju}, the magnetic sector comes directly from the dynamical modes of the original Bargmann theory. The case of electric sector is interesting as it originates from modes that are usually frozen by primary constraints in parent theory, which when deformed, drops out making them fully dynamical. Thus, the deformation process promotes the `bad' fermions into `good' ones, giving us the electric Carroll sector. This point is summarized in figure~\ref{fig:Schematic}. An additional advantage of this framework is its universal validity across  spacetime dimensions. Unlike existing constructions in the literature that depend on dimension-specific chiral projections, our approach is built upon the method of null reduction. By adapting the theory to null frames, we ensure that the Carrollian fermions are well-suited for any arbitrary spacetime dimension. 

\textbf{Note}: While this manuscript was in preparation, the preprint~\cite{Bagchi:2026bsdk} appeared exploring Carroll contraction of the relativistic Dirac action. Although their work touches upon similar themes, their analysis is based on massless chiral Weyl projections rather than strict light-cone projections. In contrast, our approach employs null reduction to systematically derive both electric and magnetic actions, along with the Carroll Clifford algebra. Moreover, our framework extends naturally to the massive case, and we analyze the energy-momentum tensor and Carroll boost invariance explicitly from a Hamiltonian perspective. Nevertheless, both works underscore the same central point: \textit{light-cone fermion dynamics are intimately connected to Carroll fermions.}
\par
This paper is organized as follows: Section 2 reviews the light-cone Dirac Lagrangian, detailing the spinor decomposition into \textit{good} and \textit{bad} fermions. Section 3 introduces the deformation to Bargmann spacetimes, demonstrating how the previously constrained `bad' fermions become dynamical. Section 4 establishes the Carroll Clifford algebra and employs null reduction to construct both the electric and magnetic Carrollian actions, followed by a canonical analysis and the computation of two-point functions for each sector. Finally, Section 5 provides concluding remarks and outlines avenues for future research.

%%%%%%%%%%%%%%%%%%%%%%%%%%%%%%%%%%%%%%%%%%%%%%
\section{Dirac Lagrangian in the light-cone formalism}

In this section, we review the Dirac Lagrangian in the light-cone formulation, beginning with the light-cone Clifford algebra. We show how the light-cone framework naturally decomposes the spinor degrees of freedom into dynamical and non-dynamical components, making the physical content of the theory more transparent. As a result, light-cone fermions have become a standard tool in quantum field theory, QCD, and particle physics~\cite{Brodsky:1997de}.
\par
The light-cone coordinates are defined as
\begin{eqnarray} \label{LCcoord}
x^+ = \frac{1}{\sqrt 2}(x^0 + x^d) \,,&& x^-= \frac{1}{\sqrt 2}(x^0 - x^d) \,,\quad x^i\,  \quad ( i,j = 1,2, ..., d-1)\,.
\end{eqnarray}
The $(d+1)$-dimensional Minkowski metric in these coordinates
\begin{equation} \label{LCds2}
dS^2= \eta^{lc}_{\mu\nu} dx^\mu dx^\nu = -2 dx^+ dx^- + \delta_{ij}dx^i dx^j\,,
\end{equation}
 has a flat Bargmann structure, as there exist two vector $\mathfrak n =\del_+$ and $\mathfrak m= \del_-$, which satisfy the property
\be
\mathfrak n^\mu \mathfrak n^\nu \eta^{lc}_{\mu\nu} = \mathfrak m^\mu \mathfrak m^\nu \eta^{lc}_{\mu\nu} = 0 \,.
\ee 
This gives rise to two Bargmann subgroups $\mathfrak{b}_{\pm}$ within the light-cone Poincar\'e algebra~\cite{Figueroa-OFarrill:2022pus}
\bea
 &\mathfrak{b}_+ : \{ P_+, P_-, P_i, M_{ij}, M_{-i}\}&\,, \\
 &\mathfrak{b}_- : \{ P_+, P_-, P_i, M_{ij}, M_{+i}\}&\,.
\eea
Each of the two Bargmann group contains within itself a codimension-one Carroll group $\mathfrak{c}_{\pm}$ and a Galilei group $\mathfrak{g}_{\pm}$. In this paper, we focus on one of these Carroll subgroups, $\mathfrak{c}_{-}$ which is spanned by 
\be
 \mathfrak{c}_- : \{ P_+, P_i, M_{ij}, M_{+i}\}\,.
\ee
 We label the algebras with the subscript $\pm$ in order to specify which null direction, $x^\pm$, is chosen as the Newtonian time. In our case, the label `$-$' indicates that $x^-$ is the Newtonian time and hence, the other null coordinate $x^+$ is taken to be Carrollian. For more details about the interesting kinematical Lie subalgebras of light-cone Poincar\'e algebra, we refer the readers to~\cite{Majumdar:2024rxg}.

%%%%%%%%%%%%%%%%%%%%%%%%%%%%%%%%%%%%%%%%%%%%%%%%%%%%
\subsection{Light-cone Clifford algebra}

The Clifford algebra associated with the metric $\eta^{lc}_{\mu\nu} $
\be \label{Clif}
\{ \G^\mu,  \G^\nu\} = 2 \eta_{lc}^{\mu\nu} \,,
\ee
also involves two null gamma matrices which satisfy
\be
(\G^+)^2 = (\G^-)^2 = 0 \,, \quad \G^+ \G^- + \G^- \G^+ = -2\,,
\ee
while for the spatial gamma matrices, we have 
\be
 (\G^i)^2 = \mathbb I\,, \quad \{ \G^i,  \G^\pm\} = 0 \,.
\ee

The presence of two null gamma matrices in the Clifford algebra allows us to define projection operators that are adapted to null frames.
\be \label{LCproj}
P^+ = -\frac{1}{2} \G^- \G^+\,, \quad P^+ = -\frac{1}{2} \G^+ \G^-\,.
\ee
These operators satisfy the following properties 
\be
(P^\pm)^2 = P^{\pm}\,, \quad P^+ + P^- = 1\,, \quad P^{-} P^+ = P^+ P^- = 0 \,.
\ee
For a Dirac spinor $\Psi$, these projectors define the two light-cone spinors 
\be \label{LC spinors}
\Psi_{(\pm)} = P^{\pm} \Psi\,, \quad \Psi = \Psi_{(+)} +\Psi_{(-)} \,, \quad (\Psi_{(\pm)}) ^\dagger = \Psi^\dagger_{(\pm)} \,.
\ee
In component form, say in four dimensions, these spinor projections read
\be \label{LCspin-comp}
 \Psi_{(+)}  = \begin{pmatrix}
\Psi_1 \\
0 \\
0 \\
\Psi_4
\end{pmatrix} \,, \quad  
\Psi_{(-)} = \begin{pmatrix}
0 \\
\Psi_2 \\
\Psi_3 \\
0
\end{pmatrix}\,.
\ee
We now consider the Lorentz transformations of the Dirac spinor
\bea
\delta \Psi = \omega^\mu{}_\nu x^\nu \frac{\del \Psi}{\del x^\mu} - \frac{1}{4} \omega^{\mu \nu} \sigma_{\mu \nu} \Psi \,,
\eea
where $\sigma^{\mu\nu} = - \frac{i}{4} [\gamma^\mu, \gamma^\nu ]$ satisfy the Lorentz algebra in light-cone coordinates. To simplify the notation, we relabel the Lorentz parameters as 
\be
\omega^{+-} = \beta\,, \quad \omega^{+i} = c^i \,, \quad \omega^{-i}= b^i \,, \quad  \omega^{ij}\,.
\ee
One finds that the light-cone spinors $\Psi_{(\pm)}$ do not mix under Lorentz transformations
\bea
\delta \Psi_{(+)} &=& \omega^\mu{}_\nu x^\nu \frac{\del \Psi_{(+)}}{\del x^\mu} - \frac{i}{8}\beta \Psi_{(+)} + \frac{i}{8} b_i \gamma^i \g^+ \Psi_{(+)} - \frac{i}{4}\omega^{ij} \sigma_{ij} \Psi_{(+)} \,, \\
\delta \Psi_{(-)} &=& \omega^\mu{}_\nu x^\nu \frac{\del \Psi_{(-)}}{\del x^\mu} + \frac{i}{8}\beta \Psi_{(-)} +  \frac{i}{8} c_i \gamma^i \g^- \Psi_{(-)} - \frac{i}{4}\omega^{ij} \sigma_{ij} \Psi_{(-)} \,.
\eea
It is worth noting that the terms involving $b_i$ and $c_i$ appear in the transformation rules of only one of the two spinors each. These correspond to the Carroll boosts associated with the two distinct Carroll subalgebras $\mathfrak c_{\pm}$ embedded within the light-cone Poincar\'{e} algebra, a point we shall elaborate on when discussing Carroll invariance of the null-reduced actions in $d$ dimensions.
\par 
Importantly, this decomposition is not merely algebraic in nature. It is intimately tied to the dynamics of the fermionic degrees of freedom, and will therefore play a central role in what follows.

%%%%%%%%%%%%%%%%%%%%%%%%%%%%%%%%%%%%%%%%%%%%%%%%%%%%

\subsection{Dirac Lagrangian: Good and bad fermions}
We now consider the light-cone Dirac Lagrangian action in $(d+1)$ dimensions\footnote{We shall use the notation $\int d^\perp x$ and $\delta^\perp (x-y)$  as a shorthand for $\int dx^1 dx^2... dx^{(d-1)}$ and $\delta^{(d-1)}(x-y)$.}

\be
S = \int dx^+ dx^- d^\perp x\, \,\mathcal L^{lc} \,,
\ee
where
\be
\mathcal L^{lc} = \overline{\Psi} (i\G^+ \del_+ + i\G^- \del_- + i\G^i \del_i - m)\Psi\,.
\ee
In terms of the light-cone spinors $\Psi_{(\pm)}$, this Lagrangian can be brought to the form
\bea \label{S-Lor}
\mathcal L^{lc}&=&-i \sqrt 2 \Psi^\dagger_{(+)} \del_+ \Psi_{(+)}  -i \sqrt 2 \Psi^\dagger_{(-)} \del_- \Psi_{(-)}  \\
&&+  \frac{i}{\sqrt 2} \left( \Psi^\dagger_{(-)} \G^+ \G^i \del_i \Psi_{(+)} +  \Psi^\dagger_{(+)} \G^- \G^i \del_i \Psi_{(-)} \right)  - \frac{m}{\sqrt 2} \left( \Psi^\dagger_{(-)} \G^+  \Psi_{(+)} +  \Psi^\dagger_{(+)} \G^-\Psi_{(-)} \right) \,. \nn
\eea
There is no $\del_+ \Psi_{(-)}$ term in the action indicating that $ \Psi_{(-)}$ is not dynamical. This points to the presence of an extra primary constraint in the light-cone theory, which becomes more transparent from the conjugate momenta
\bea
\Pi_{(+)} &=& \frac{\delta \mathcal L}{\delta (\del_+ \Psi_{(+)})} = -i \sqrt 2 \Psi^\dagger_{(+)} \,, \\
\Pi_{(-)} &=& \frac{\delta \mathcal L}{\delta (\del_+ \Psi_{(-)})} =0  \,.
\eea
The first expression corresponds to the usual second-class constraint of fermions, which allows one to eliminate the momenta and work with Dirac brackets, $\{ \Psi_{(+)} (x), \Psi^\dagger_{(+)} (y)\}$, rather than Poisson brackets, $\{ \Psi_{(+)} (x), \Pi_{(+)} (y)\}$. In the light-cone literature, the $\Psi_{(+)}$ projection is accordingly called the `good' fermion, as it captures the dynamics of the free Dirac theory.
\par 
The second expression, by contrast, amounts to a first-class primary constraint \textit{unique to light-cone spinors}. This gauge constraint renders $\Psi_{(-)}$ non-dynamical; it can be gauged away via $\Psi_{(-)} \rightarrow \Psi_{(-)} + \chi$, where $\chi$ is an arbitrary spinor parameter. The $\Psi_{(-)}$ projection is therefore called the `bad' fermion, as it can be eliminated through its own equation of motion. This is indeed the standard practice in light-cone QED and QCD, where gauge-fixing is employed to remove the non-dynamical component~\cite{Kogut:1969xa,Bjorken:1970ah,Brodsky:1997de}. However, for our purposes, we retain both $\Psi_{(\pm)}$ projections, as we are interested in the most general Carroll fermion theories accessible via null reduction.
\par
For bosonic theories, the light-cone action with similar primary constraints yields magnetic Carroll theories upon null reduction~\cite{Majumdar:2025ju}. To access the electric sector, one must instead deform the light-cone action so as to remove this constraint -- resulting in what we call the deformed light-cone Bargmann action. The deformation takes us from a light-cone Minkowski background to the more general Bargmann geometry.
\par
As discussed in section~\ref{Carr-sec} below, the $\Psi_{(+)}$ projection corresponds to the magnetic Carroll sector upon null reduction, while $\Psi_{(-)}$, which becomes dynamical on the Bargmann background, yields the electric Carroll sector upon null reduction.

%%%%%%%%%%%%%%%%%%%%%%%%%%%%%%%%%%%%%%%%%%%%%%%%%%%%

\subsection{Light-cone projections versus chiral projections}
Before we turn to Carroll fermions, we first focus on some key distinctions between the light-cone projectors and the commonly used Weyl or chiral projectors. As alluded to above, the light-cone projectors $P^\pm$ separates out the dynamical and non-dynamical components of a spinor.
\par
On the other hand, the chiral projectors defines left and right-handed spinors. In even dimensions $d+1 = 2n$, there exists a chirality matrix $\G^*$ matrix (or $\G^5$ in four dimensions) 
\be
\G^* = i^{n-1} \G^0 \G^1 \ldots \G^{d}\,.
\ee
which allows us to decompose a Dirac spinor into chiral projections. Without loss of generality, we restrict our discussion to four dimensions in this subsection, such that the chirality matrix is given by
\be
\G^5 = i \G^0 \G^1 \G^2 \G^{3} \,.
\ee
The $\G^5$ anticommutes with all the $\G^\mu$ and satisfies $(\G^5)^2 = \mathbb I$. The chiral projection operators then read
\be \label{LRproj}
P_{L(R)} = \frac{1}{2} (1 \pm i\G^5)\,.
\ee
The left- and right-handed spinors are defined as
\be
 \psi_L= P_L \psi   \,, \quad  
\psi_R=  P_R \psi \,,
\ee
such that the four-component Dirac spinors $\Psi$ splits into
\be
\Psi = \begin{pmatrix}
\Psi_L \\
\Psi_R
\end{pmatrix}\,,
\ee
where the $\Psi_{L}$ and $\Psi_{R}$ are the  two-component Weyl spinors 
\be
 \quad \Psi_L =  \begin{pmatrix}
\Psi_1 \\
\Psi_2
\end{pmatrix} \,, \quad \Psi_R =\begin{pmatrix}
\Psi_3 \\
\Psi_4
\end{pmatrix}\,.
\ee
An important distinction concerns the dimensional validity of each projection.  The chiral projections $\Psi_{L/R}$ are eigenstates of $\Gamma^5$ in four dimensions (or $\Gamma^*$ in any even dimension),
\be
\Gamma^5 \Psi_{L} = +\Psi_{L}\,, \qquad \Gamma^5 \Psi_{R} = -\Psi_{R}\,,
\ee
and as such, they are only meaningful in even dimensions, where $\Gamma^*$ is non-trivial. In odd dimensions, $\Gamma^*$ reduces to a multiple of the identity and can no longer distinguish chirality. The light-cone projections, by contrast, are defined through the null gamma matrices $\Gamma^\pm$, which are well-defined in any dimension. Consequently, the light-cone spinors $\Psi_{(\pm)}$ exist in any arbitrary dimensions, making the light-cone formalism more general.
\par
A second distinction concerns the structure of the resulting subspaces. The chiral projectors split the four-component Dirac spinor into two two-component Weyl spinors, whereas the light-cone projectors split it into two four-component spinors, as shown in \eqref{LCspin-comp}. In both cases, one obtains two invariant subspaces spanned by $\Psi_{L/R}$ or $\Psi_{(\pm)}$ respectively. However, the nature of these 
decompositions differs fundamentally. The chiral projections yield a fully  decomposable representation, with each Weyl spinor furnishing an irreducible representation of the Lorentz group. The light-cone splitting, on the other hand, produces a representation that is reducible but indecomposable. 
\par
The mass term in the Dirac Lagrangian explicitly breaks chiral symmetry, mixing $\Psi_L$ and $\Psi_R$. Consequently, Carroll constructions appearing in the existing literature that rely on chiral projections are mostly restricted to the massless case~\cite{Bagchi:2022eui, Koutrolikos:2023evq, Ara:2024fbr, Ekiz:2025hdn, Bagchi:2026bsdk}. The light-cone approach does not suffer from this limitation, making it the more natural framework for studying both massless and massive Carroll fermions.

%%%%%%%%%%%%%%%%%%%%%%%%%%%%%%%%%%%%%%%%%%%%%%%%%%%%

\section{Deformation to Bargmann spacetimes}
The key ingredient of our construction relies on the work of~\cite{Duval:2014uva}, where it was shown how Carrollian theories can be derived from a higher-dimensional parent theory in a Bargmann spacetime. The light-cone Minkowski metric belongs to a special class of Bargmann spacetimes, due to the presence of not one but two covariantly constant null vectors. However, as we found in~\cite{Majumdar:2025ju} for bosonic theories, this turns out to be an obstacle rather than an advantage: in order to obtain \textit{both} the electric and magnetic Carroll actions via null reduction from a \textit{single} parent action, one must deform the light-cone Lorentzian metric to a more general Bargmann metric
\be \label{Barg-metric}
G_{\mu \nu} =\begin{pmatrix}
0&-1 & 0 \\
-1&-\alpha& 0 \\
0 & 0 & \delta_{ij}
\end{pmatrix} \,,
\ee
where $\alpha$ is a constant deformation parameter relating the Lorentzian light-cone coordinates to the Bargmann light-cone coordinates~\cite{Lenz:1991sa, Burkardt:1995ct}
\bea \label{Lenz-coord}
x^+ \to x^+ + \frac{\alpha}{2}\, x^- \,,  \quad x^-\to x^- \,,  \quad x^i\to x^i \,.
\eea
We now show how both electric and magnetic Carroll fermion theories are obtained via null reduction of the Dirac Lagrangian on this general Bargmann spacetime.
\subsection{Light-cone Bargmann spinors}
On the Bargmann spacetime, the Clifford algebra is simply a deformation of the light-cone Lorentzian Clifford algebra
\be \label{Barg-Cliff}
\{ \G^\mu\,, \G^\nu \}= 2G^{\mu \nu} \,,
\ee
with $(\G^-)^2 =0$, $(\G^i)^2 =\mathbb I$ as before, but $(\G^+)^2 = \alpha$.
The light-cone projectors have the same form
\be
\hat{P^+} = -\frac{1}{2} \G^- \G^+\,, \quad \hat{P^-} = -\frac{1}{2} \G^+ \G^-\,, \quad 
\ee
and satisfy 
\be
(\hat{P^\pm})^2 = \hat{P^\pm}\,, \quad \hat{P^\pm}\hat{P^\mp} =0\,, \quad  \hat{P^+}+\hat{P^-} = 1 \,.
\ee 
However, the $\G^\mu$ now satisfy the Bargmann Clifford algebra \eqref{Barg-Cliff}. We can now define Bargmann light-cone spinors as follows
\be
\Psi_{(+)} = \hat{P^+}  \Psi \,, \quad \Psi_{(-)} = \hat{P^-}  \Psi \,.
\ee
Therefore, just like in the Lorentzian case, a Dirac fermion in a Bargmann spacetime splits into `good' and `bad' parts, $\Psi_{(+)}$ and $\Psi_{(-)}$ respectively. 

\subsection{When bad fermions become good}
We now consider the Dirac Lagrangian in $(d+1)$-dimensional light-cone Bargmann spacetime
\be
\mathcal S_\alpha = \int dx^+ dx^- d^\perp x \, \bar{\Psi} (i G_{\mu \nu} \G^\mu \partial^\nu - m) \Psi \,.
\ee
Using the properties of the projection operators $\hat{P}^{\pm}$ and other $\hat{\Gamma}^\mu$ identities, the above action can be brought to the form
\bea \label{S-Barg}
S^{Barg}_\alpha &=&  \int dx^+ dx^- d^\perp x\, \bigg[ -\frac{i}{\sqrt 2}\alpha\,  \Psi^\dagger_{(-)} \del_+\Psi_{(-)}  -i \sqrt 2 \Psi^\dagger_{(+)} \del_+ \Psi_{(+)}  -i \sqrt 2 \Psi^\dagger_{(-)} \del_- \Psi_{(-)}  \\
&+&  \frac{i}{\sqrt 2} \left( \Psi^\dagger_{(-)} \G^+ \G^i \del_i \Psi_{(+)} +  \Psi^\dagger_{(+)} \G^- \G^i \del_i \Psi_{(-)} \right)  - \frac{m}{\sqrt 2} \left( \Psi^\dagger_{(-)} \G^+  \Psi_{(+)} +  \Psi^\dagger_{(+)} \G^-\Psi_{(-)} \right) \bigg]\,. \nn
\eea
Crucially, this action differs from the Lorentzian action in \eqref{S-Lor} by a single $\alpha$-dependent term in the kinetic part, that makes $\Psi_{(-)}$ dynamical, effectively turning the `bad' fermion to the into `good'. The fact that the $\Psi_{(-)}$ field is not constrained anymore is best reflected by its conjugate momentum
\bea
\Pi_{(-)} &=& -\frac{i}{\sqrt 2}  \Psi^\dagger_{(-)} \,,
\eea
which does not amount to a first-class primary constraint, unlike in the Lorentzian case.
\par Therefore, the light-cone projections on the Bargmann background no longer isolate dynamical from non-dynamical components of the spinor. Nevertheless, the decomposition into $\Psi_{(+)}$ and $\Psi_{(-)}$ continues to serve as a useful organizing principle, distinguishing between the magnetic and electric Carroll sectors respectively, as we shall show below.

%%%%%%%%%%%%%%%%%%%%%%%%%%%%%%%%%%%%%%%%%%%%%%%%%%%%

\section{Carroll fermions} \label{Carr-sec}

\subsection{Carroll Clifford algebra}
\begin{figure}[h]
    \centering
    \vskip 0.5cm
    \includegraphics[width=0.9\textwidth]{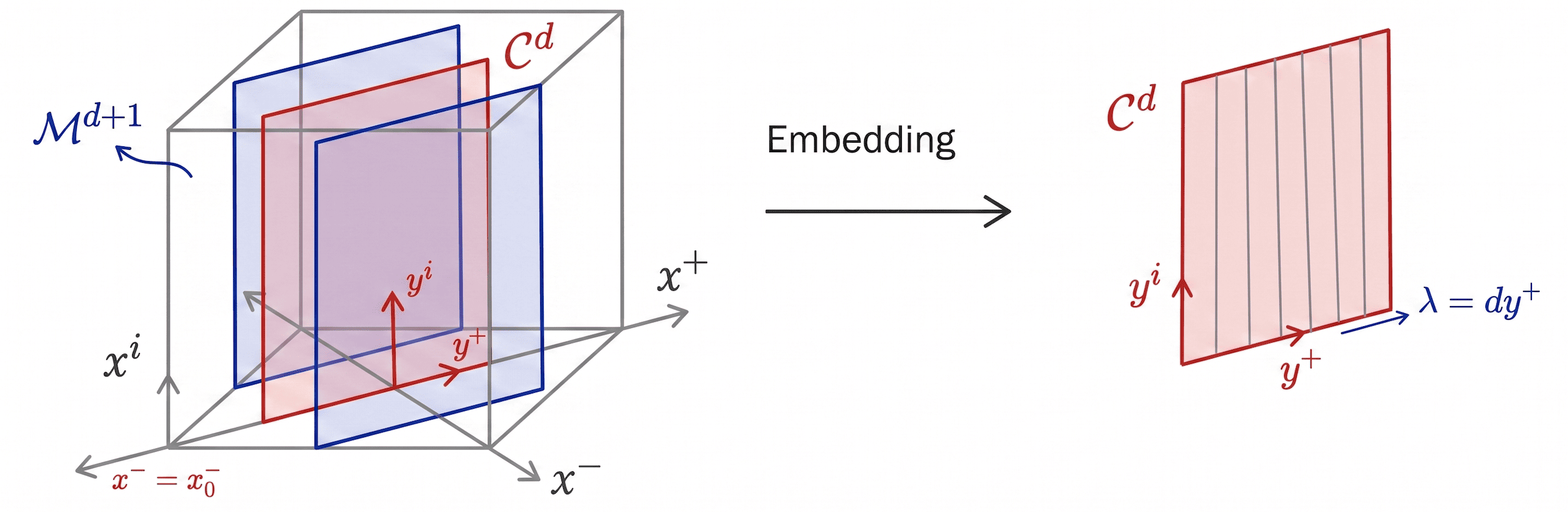}
    \caption{Carroll geometry from null reduction: A Carroll manifold $(C^d, g_{ab}, \mathfrak n^a)$ embedded a Bargmann manifold endowed with the light-cone Minkwoski metric $(\mathcal M^{d+1}, \eta^{l.c.}_{\mu\nu}, \mathfrak n^\mu)$}
    \label{fig:embed}
        \vskip 0.5cm
\end{figure}

In order to obtain the Carroll Clifford algebra from the relativistic case, we first discuss the Carrollian geometry of a null hypersurface $\Sigma$ embedded into a Bargmann manifold, as depicted in figure~\ref{fig:embed}. 
\par Let us consider a $(d+1)$-dimensional Bargmann manifold $M^{d+1}$ with the metric $G_{\mu\nu}$ as in \eqref{Barg-metric}, where $x^\mu = (x^+,x^-,x^i)$ are the local coordinates on $M$. The vector $\mathfrak n = \partial_+$, which satisfies
\be
 \mathfrak  n^\mu  \mathfrak  n^\nu G_{\mu\nu} = 0\,,
 \ee
 is the preferred null direction. The associated clock 1-form is given by 
\be
\tau_\mu = G_{\mu \nu} \mathfrak n^\mu \quad \Rightarrow \quad \tau = -dx^-\,.
\ee
One finds that ker $\tau$ defines a foliation on $M$. Each leaf of this foliation, say $\Sigma: x^- =x^-_0$, is endowed with a Carroll structure. The light front $\Sigma$ at $x^- = x^-_0$, where $x^-_0$ is a constant, may be viewed as the embedding $\Phi : \Sigma \hookrightarrow M$ given by
\be
x^+(y) = y^+\,, \quad x^-(y) = x^-_0\,, \quad x^i(y) = y^i\,,
\ee
where $y^a = (y^+, y^i)$ are the local coordinates on $\Sigma^d$. This hypersurface $\Sigma$ us endowed with a degenerate metric 
\be
g_{ab} = \begin{pmatrix}
0 & 0 \\
0  & \delta_{ij}
\end{pmatrix} \,, 
\ee
and a vector field $\mathfrak n^a = \del_+$, which lies in the kernel of $g_{ab}$ such that $\mathfrak n^a g_{ab} = 0$. Thus, we see that the light front indeed has a flat Carrollian structure $(\Sigma^d, g_{ab}, \mathfrak n^a)$. 

\par
The isometries of the Carroll manifold  given by
\be
L_\xi \mathfrak n^a = 0 \,, \quad  L_\xi g_{ab} =0 \quad  \Rightarrow \quad  \xi = \xi^+ (y^i) \del_+ + (\omega^i_j y^j + a^i) \del_i\,.
\ee
Moreover, if we restrict $\xi$ to be linear in the coordinates, we get the finite-dimensional Carroll group spanned by $\mathfrak c_{-}$
\be \label{Carroll}
\xi^+ = b_i y^i+ a^+\,, \quad \xi^i = \omega^i_j y^j + a^i\,, 
\ee
where $b_i$ are the Carroll boosts, $\omega_{ij}$ the spatial rotations and $(a^+, a^i)$ the spacetime translations.
\par
In order to define a `pseudo-inverse' of $g_{ab}$, we consider at every point on $\Sigma$,  a second null vector $\lambda^\mu$, not parallel to $ \mathfrak n^\mu$ such that
\be
G_{\mu \nu} \lambda^\mu \lambda^\nu = 0\,, \quad G_{\mu \nu} n^\mu \lambda^\mu = 1\,.
\ee
The vector $\lambda^\mu$ is called the rigging of $\Sigma$ and, in the present case, reads 
\be
\lambda = -\frac{\alpha}{2}\del_+ - \del_-\,.
\ee
 The rigging vector forms a part of the extrinsic Carroll data, which has been shown to play an important role in null geometries (see~\cite{Ciambelli:2025unn} for more details). We can now define a 1-form on $\Sigma$ as 
\be
{\lambda_a = G_{\mu \nu} \lambda^\mu e^\nu_a\,, \quad \lambda_a\mathfrak n^a  = 1 \quad \Rightarrow \quad \lambda = dy^+}\,.
\ee
Using $\lambda_a$ we can find a symmetric tensor $g^{ab}$ with the properties
\be \label{inv-g}
g_{ab}\,g^{bc} = \delta_a{}^c - \lambda_a \mathfrak n^c \,, \quad \lambda_a g^{ab} = 0 \,,
\ee
which has the form
\be
 g^{ab} = \begin{pmatrix}
0 & 0 \\
0  & \delta^{ij}
\end{pmatrix} \,.
\ee
On the Bargmann manifold $\mathcal{M}$, the Clifford algebra takes the form
\be
\{ \Gamma^\mu, \Gamma^\nu \} = 2G^{\mu \nu}\,,
\ee
or equivalently, $\{ \Gamma_\mu, \Gamma_\nu \} = 2G_{\mu \nu}$. The gamma matrices on the Carroll manifold $\Sigma$ are obtained by pulling back $\Gamma^\mu$ to the null hypersurface 
\be
\gamma^a = e^{a}{}_{\mu}\, \Gamma^\mu\,,
\ee
such that they satisfy the Carroll Clifford algebra
\be
\{ \gamma^a, \gamma^b\} = 2g^{ab}\,,
\ee
where $g^{ab}$ is the `pseudo-inverse' of the degenerate metric on $\Sigma$, satisfying the relation~\eqref{inv-g}. Equivalently, lowering the indices with the Carroll metric $g_{ab}$ yields
\be
\{ \gamma_a, \gamma_b\} = 2g_{ab}\,,
\ee
where
\be
\gamma_a = G_{\mu\nu}\,\Gamma^\mu\, e^\nu{}_a \,.
\ee
In particular, this implies $\gamma_+ = \Gamma^-$, and thus $(\gamma^+)^\dagger = \gamma_+$. This  ensures the hermiticity of the resulting Carroll actions. The relation $\{\Gamma^+, \Gamma^-\} = -2$, while no longer part of the intrinsic Carroll Clifford algebra, survives as an algebraic relation on $\Sigma$ of the form
\be
\gamma^+\gamma_+ + \gamma_+\gamma^+ = -2\,.
\ee
The appearance of $\Gamma^-$ through $\gamma_+$ may be interpreted as part of the extrinsic spinor data associated with the embedding of the Carroll manifold in the ambient Bargmann space. Indices on the Carrollian gamma matrices may be freely raised and lowered using the appropriate geometric objects on $\Sigma$.

%%%%%%%%%%%%%%%%%%%%%%%%%%%%%%%%%%%%%%%%%%%%%%%%%%%%
\subsection{Carrollian action from null reduction}

In this subsection, we describe a general procedure to restrict a Lorentzian field theory onto a null hyperplane using a simple trick that allows us to directly extract the Carrollian action from the higher-dimensional parent action~\cite{Chen:2023pqf}
\par
We consider an action ${}^{(d+1)}\mathcal S$ in light-cone Bargmann spacetime for a theory involving a  collection of fermionic fields $\{\Psi\}$ and their conjugate momenta $\{\Pi_\Psi\}$, which are related to $\{\Psi^\dagger\}$ through some primary constraints
\be \label{GenS}
{}^{(d+1)}\mathcal S[\Psi, \Psi^\dagger] = \int dx^+ dx^- d^\perp x\ {}^{(d+1)} \mathcal L_{Barg} [\Psi, \Psi^\dagger]\,.
\ee
The canonical Poisson brackets and the symplectic 2-form are thus given by
\bea
&&\{ \Psi (x), \Pi_{\Psi} (y) \} ~=~ \delta (x^- -y^-) \delta^\perp(x-y)\,,\\[0.1cm]
 && {}^{(d+1)} \Omega~ =~ \int dx^- d^\perp x \, \,d_V \Pi_{\Psi} \wedge d_V \Psi\,,
\eea
where $d_V$ denotes exterior derivative in phase space.
\par
\begin{figure}[h]
\centering
\includegraphics[width=2.5in]{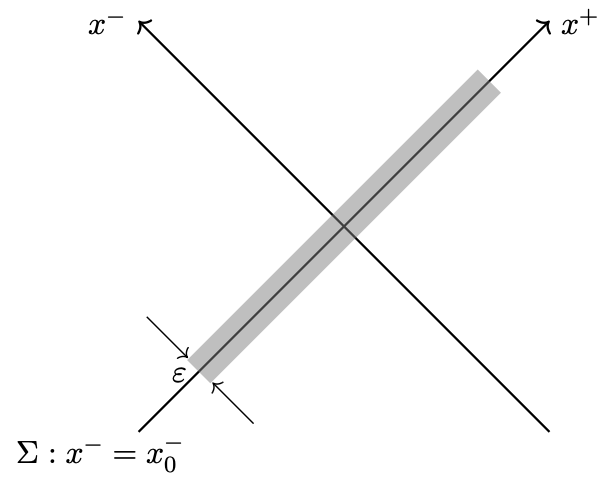}
\caption{Null reduction to the hypersurface $\Sigma: x^- = x^-_0$}
\label{fig:hypersurface}
\end{figure}

 
To evaluate the effective action restricted to any $\Sigma: x^- = x_0^- $ hypersurface, we introduce a Dirac delta distribution with a smearing function~\cite{Chen:2023pqf}, as shown in Figure~\ref{fig:hypersurface}. 
\bea
\delta_\varepsilon (x^- -x^-_0) &=& 
\begin{dcases}
      \frac{1}{\varepsilon} \,, &x^-_0 - \frac{\varepsilon}{2} < x^- < x^-_0 + \frac{\varepsilon}{2}  \\
0 \,,& \text{otherwise} 
    \end{dcases}
\eea
such that 
\be
\lim_{\varepsilon \to 0} \delta_\varepsilon (x^- -x^-_0) = \delta (x^- -x^-_0) \,. 
\ee

By introducing the smearing function in \eqref{GenS}, we define a family of actions $\mathcal S_\varepsilon$ such that, in the limit $\varepsilon \to 0$, we obtain the Carrollian action
\bea
{}^{(d)}\mathcal S_{Carr} ~\equiv~ \lim_{\varepsilon \to 0} {}^{(d+1)}\mathcal S_{\varepsilon} \,.
\eea
To obtain a finite, non-vanishing result, one must specify the behavior of the $(d+1)$-dimensional fields near $x^- = x^-_0$, ensuring that the $\varepsilon \to 0$ limit yields a well-defined Carrollian action. As we shall see, there are precisely two behaviors of the higher-dimensional fields near a constant $x^-$ plane that preserve the canonical structure under null reduction,
\be
{}^{(d)}\Omega_{Carr} ~\equiv~ \lim_{\varepsilon \to 0} {}^{(d+1)}\Omega_{\varepsilon}\,,
\ee
corresponding respectively to the magnetic and electric Carroll theories in one lower dimension. Henceforth, we specialize to $3+1$ dimensions for concreteness. However, all constructions and results presented below hold in arbitrary dimensions.
\par
Starting with Bargmann-invariant Dirac action in \eqref{S-Barg} in four dimensions, we now restrict the theory on a null hypersurface in a way that preserves the form of  $\Omega_{Barg}$ in the limit $\varepsilon \to 0$
\bea
\Omega_{Barg} &=& \int dx^- d^\perp x \left( - \frac{i}{\sqrt 2} \alpha\, d_V\Psi_{(-)}^\dagger\wedge d_V\Psi_{(-)} -i \sqrt 2\,  d_V\Psi_{(+)}^\dagger\wedge d_V\Psi_{(+)} \right) \,.
\eea
In the vicinity of the $x^- = x^-_0$ plane, we find that there are two possible ways to map the Bargmann fields $\Psi_{(\pm)}$ to Carrollian fields $\psi_{(\pm)}$ as follows
\begin{itemize}
\item \textit{Case I:}
 \be \label{elec-map}
\Psi_{(+)} \rightarrow \varepsilon \psi_{(+)}\,, \quad \Psi_{(+)} \rightarrow \varepsilon \psi_{(+)}\,, \quad  \alpha \rightarrow  \frac{\alpha}{\varepsilon^2} \,.
\ee
This choice corresponds to the electric sector with the Carroll symplectic form
\be
\Omega^{elec}_{Carr} = - \frac{i}{\sqrt 2} \alpha  \int d^\perp x \ d_V\psi_{(-)}^\dagger\wedge d_V\psi_{(-)} \,.
\ee
\item \textit{Case II:}
\be \label{mag-map}
\Psi_{(+)} \rightarrow  \psi_{(+)}\, \quad \Psi_{(+)} \rightarrow \psi_{(+)}\,, \quad \alpha \rightarrow \varepsilon \alpha\, .
\ee
This choice corresponds to the magnetic sector with the Carroll symplectic form
\be
\Omega^{mag}_{Carr} = -i\sqrt 2 \alpha  \int d^\perp x \, d_V\psi_{(+)}^\dagger\wedge d_V\psi_{(+)} \,.
\ee
\end{itemize}
In the following sections, we will show how the Carrollian theories arising from these rescalings correspond to the electric and magnetic cases. 
\par 
Note that the rescaling $\alpha \to \varepsilon \alpha$ is consistent with the fact that the magnetic sector corresponds to $\alpha = 0$. This is because one can obtain the magnetic action directly by null reduction the from the Lorentzian Dirac action. But the electric sector is only accessible for $\alpha \neq 0$, indicating why it was essential to consider the $\alpha$-deformation to the Bargmann geometry, as depicted schematically in Figure \ref{fig:Schematic}.

%%%%%%%%%%%%%%%%%%%%%%%%%%%%%%%%%%%%%%%%%%%%%%%%%%%%

\subsection{Electric branch}
We consider the following map from the Bargmann fields $\Psi_{(\pm)}$ to Carrollian fields $\psi_{(\pm)}$ 
\be
\Psi_{(+)} \rightarrow \varepsilon \psi_{(+)}\,, \quad \Psi_{(+)} \rightarrow \varepsilon \psi_{(+)}\,, \quad  \alpha \rightarrow  \frac{\alpha}{\varepsilon^2} \,, \quad m \rightarrow  \frac{m}{\varepsilon^2} \,,
\ee
where we have also rescaled the parameters $\alpha$ and $m$. In the limit $\varepsilon \to 0$, the action becomes
\bea
&&\mathcal S^{elec}_{Carr}= \int dx^+ d^\perp x \, \mathcal L^{elec}  [\psi_{(-)}] \,,
\eea
where the electric Lagrangian reads
\bea
&& \mathcal L^{elec} =-\frac{i}{\sqrt 2} \alpha \psi^\dagger_{(-)} \del_+ \psi_{(-)} - \frac{m}{\sqrt 2} \left(\psi^\dagger_{(-)} \g^+ \psi_{(+)} + \psi^\dagger_{(+)} \g_+ \psi_{(-)} \right)  \,.
\eea
The above action has an electric Carroll structure as it only involves time derivatives. On the Carroll manifold, $\psi_{(+)}$ and $\psi_{(-)}$ fields can be related through a \textit{null constraint} $\g^+ \psi_{(+)} = \psi_{(-)}$, which allows us to eliminate $\psi_{(+)}$ and express the electric Carroll action in terms of $\psi_{(-)}$ alone. Finally, the electric Carroll Lagrangian takes the form\footnote{For convenience, we have relabeled the fields as $\psi_{(-)} \to (2/\alpha)^{1/2} \psi_{(-)}$ and the mass $m \to m \alpha /2$.}
\be \label{L-elec}
\mathcal L^{elec}= - i\sqrt 2 \psi^\dagger_{(-)} \del_+ \psi_{(-)} - \sqrt 2 m\,  \psi^\dagger_{(-)} \psi_{(-)}  \,.
\ee
This matches the form of the electric Carroll fermion action found in~\cite{Bergshoeff:2023vfd}.

%%%%%%%%%%%%%%%%%%%%%%%%%%%%%%%%%%%%%%%%%%%%%%%%%%%%

\subsubsection{Canonical analysis}
In the canonical formulation, the necessary and sufficient conditions for a Hamiltonian theory to be Carrollian are as follows~\cite{Henneaux:2021yzg}
\begin{enumerate}
\item The energy and momentum densities, $\mathcal H^C$ and $\mathcal P^C_i$ satisfy the Carroll commutation relations
\bea \label{H-P Comm}
\left[\mathcal{H}^{C}(x),\, \mathcal{H}^{C}(y) \right] &=& 0  \label{HH} \,, \\
\left[\mathcal{H}^{C}(x),\, \mathcal P^{C}_i (y)\right] &= &  \mathcal H^{C}(y) \,  \, \del_i \delta^\perp (x-y) \,, \\
\left[ \mathcal P^{C}_i (x),\, \mathcal P^{C}_j(y) \right] &=& \frac{1}{2}  \big[ \del_j \delta^\perp (x-y) \mathcal P^C_i (y) + \del_i \delta^\perp (x-y) \mathcal P^C_j (x) \big]. \label{PP}
\eea
\item The action is invariant under the Carroll transformations generated by
 \be \label{G-Carr}
G [{\xi^+, \xi^i}] = \int d^\perp x \ (\xi^+ \mathcal H^{C} + \xi^i \mathcal P^{C}_i)\,,
\ee
where $\xi^+, \xi^i$ are listed in \eqref{Carroll}.
\end{enumerate}
We will show that the electric action discussed above satisfies both the properties.
\par
We first note that the conjugate momenta in this case reads 
$ \pi_{(-)} = - i {\sqrt 2} \psi^\dagger_{(-)}\, $,
which implies the Poisson bracket, or appropriately the Dirac bracket takes the simple form
\be
\{ \psi_{\alpha(-)} (x), \psi^\dagger_{\beta(-)} (y)\} \Big{|}_{x^+ = y^+} = \frac{i}{\sqrt 2} \, \delta_{\alpha \beta} \delta^\perp (x-y)  \,,
\ee
where $\alpha, \beta$ denote the spinor indices. The canonical Hamiltonian density is
\be
\mathcal H^{elec} = \sqrt 2 m\, \psi^\dagger_{(-)}\psi_{(-)}\,,
\ee
which vanishes for massless fields. With the momentum density defined as 
\be
\mathcal P^{elec}_i =  - i {\sqrt 2} \psi^\dagger_{(-)} \del_i \psi_{(-)}\,,
\ee
one can easily verify that the commutation relations \eqref{H-P Comm} - \eqref{PP} are satisfied. 
\par
It also follows that the the Carroll transformations generated by $G [{\xi^+, \xi^i}] $
\be \label{elec-boost}
\delta \psi_{(-)} = \xi^+ \frac{\del \psi_{(-)}}{\del x^+} + \xi^i \frac{\del \psi_{(-)}}{\del x^i} + \frac{i}{8} c_i \g^i \gamma_+ \psi_{(-)} - \frac{i}{4} \omega^{ij} \sigma_{ij} \psi_{(-)}\,,
\ee
render the electric Carroll action \eqref{L-elec} invariant. 

 %%%%%%%%%%%%%%%%%%%%%%%%%%%%%%%%%%%%%%%%%%%%%%%%%%%%

\subsubsection{Two-point functions}
The equation of motion obtained from the electric action is
\be \label{elecEOM}
 (i\del_+ +  m)\, {\psi_{(-)}} = 0 \,,
\ee
whose most general solution has the form
\be \label{elec-sol}
\psi_{(-)} (x^+, x^i) = \psi^0_{(-)} (x^i)\,  e^{imx^+}\,.
\ee
This solution also admits a smooth limit to the massless case, $\psi_{(-)} ^{m=0}(x^+, x^i) = \psi^0_{(-)}(x^i)$.
 Using the Green's function method
\be
(i\del_+ + m) G_{\mathrm{elec}} (x, y) = \delta(x^+ - y^+) \delta^\perp (x-y)\,,
\ee
the two-point function is found to be
\be
G_{\mathrm{elec}}(x^+-y^+, \mathbf{x}_\perp-\mathbf{y}_\perp) = -ie^{im(x^+-y^+)} \Theta(x^+-y^+) \delta^\perp (\mathbf{x}_\perp-\mathbf{y}_\perp)\,,
\ee
in agreement with the existing literature~\cite{Sharma:2025rug} and admits a smooth massless limit. \par
The electric theory may also be quantized via standard canonical methods~\cite{Sharma:2025rug,Ekiz:2025hdn}. Since all $x^+$ dependence is carried by the phase $e^{imx^+}$,  a mode expansion is performed only over the transverse spatial directions,
\bea
\psi_{(-)\alpha} (x^+, \mathbf{x}_\perp) = \int \frac{d^2k}{2\pi} e^{imx^+} e^{i \mathbf{k}\cdot \mathbf{x_\perp}} b_{\alpha} (\mathbf{k}) \,,\\
\psi_{(-)\alpha}^\dagger (x^+, \mathbf{x}_\perp) = \int \frac{d^2k}{2\pi} e^{-imx^+} e^{i \mathbf{k}\cdot \mathbf{x_\perp}} b^\dagger_{\alpha} (\mathbf{k}) \,.
\eea
The operators $b_{\alpha} (\mathbf{k}) $ and $b^\dagger_{\alpha} (\mathbf{k}) $ are labelled by the transverse momentum $\mathbf{k}$. The only non-vanishing anticommutator is
\be
\{ b_\alpha (\mathbf{k}), b^\dagger_\beta (\mathbf{k'})\} = \frac{1}{\sqrt 2}(2\pi) \delta_{\alpha \beta} \delta^\perp (\mathbf{k}- \mathbf{k'})\,.
\ee 
The Carroll Hamiltonian expressed in terms of these operators takes the familiar form
\be
H^{elec} = \sqrt 2 m \int d^2k\, b^\dagger_{\alpha} (\mathbf{k}) b_{\alpha} (\mathbf{k}) \,,
\ee
so that each transverse mode carries the same energy $\sqrt 2m$. The electric carroll vacuum is defined as $ b_{\alpha} (\mathbf{k})|0\rangle = 0$ and the single-particle states are created by acting with $ b^\dagger_{\alpha} (\mathbf{k})$. As expected, the excitations do not propagate which is a signature of Carrollian ultralocality.

 %%%%%%%%%%%%%%%%%%%%%%%%%%%%%%%%%%%%%%%%%%%%%%%%%%%%
\subsection{Magnetic branch}
We now turn to the second choice for mapping the Bargmann fields to the Carrollian fields $\psi_{(\pm)}$, which preserves the form of $\Omega_{Barg}$, as in  \eqref{mag-map}
\be
\Psi_{(+)} \rightarrow  \psi_{(+)}\,, \quad \Psi_{(+)} \rightarrow \psi_{(+)}\,, \quad \alpha \rightarrow \varepsilon \alpha\,.
\ee
In the limit $\varepsilon \to 0$, the action reduces to 
\bea
&&\mathcal S^{M}_{Carr} = \int d^3x\, \mathcal L^M  [\psi_{(+)}, \psi_{(-)}] \,,
\eea
with the magnetic Lagrangian 
\bea
&&\mathcal L^M= -i\sqrt 2  \psi^\dagger_{(+)} \del_+ \psi_{(+)} + \frac{1}{\sqrt 2}  \psi^\dagger_{(-)} \g^+ (i\g^i \del_i -m)\psi_{(+)} +\frac{1}{\sqrt 2}  \psi^\dagger_{(+)} \g_+  (i\g^i \del_i -m)\psi_{(-)} \,.
\eea
 We can now impose a different null constraint, $\g_+ \psi_{(-)} = \psi_{(+)}$, which after some manipulations, reduces the magnetic Carroll Lagrangian to the form
\be \label{L-mag}
\mathcal L^{M} [\psi_{(+)}, \psi_{(-)}]= -\frac{i}{\sqrt 2} \left(\psi^\dagger_{(-)} \g^+ \del_+ \psi_{(+)} + \psi^\dagger_{(+)} \g^+ \del_+ \psi_{(-)} \right)+ \sqrt 2 \psi^\dagger_{(+)}   (i\g^i \del_i -m)\psi_{(+)}  \,.
\ee
The key difference from the electric case is that the magnetic action cannot be expressed solely in terms of the physical modes $\psi_{(+)}$. This is in agreement with magnetic Carroll actions for fermions that were found in~\cite{Bergshoeff:2023vfd, Mele:2023lhp}. The  $\psi_{(-)}$ field acts as a Lagrange multiplier, which enforce the constraint $\partial_+ \psi_{(+)} = 0$, and hence cannot be eliminated from the phase space. As a consequence, one must work with the enlarged phase space $\{\psi_{(\pm)}, \psi^\dagger_{(\pm)}\}$, a fact that has important implications for the quantization of the theory.

%%%%%%%%%%%%%%%%%%%%%%%%%%%%%%%%%%%%%%%%%%%%%%%%%%%%

\subsubsection{Canonical analysis}
The conjugate momenta for $\psi_{(+)}$ and $\psi_{(-)}$ read
\be
 \pi_{(+)} = - \frac{i}{\sqrt 2} \psi^\dagger_{(-)} \g^+\,,\quad \pi_{(-)} = - \frac{i}{\sqrt 2} \psi^\dagger_{(+)} \g_+
 \,.
\ee 
The kinetic structure is off-diagonal, with $\psi_{(+)}$ conjugate to $\psi_{(-)}$ and vice-versa.  So the theory carries two coupled second-class fermionic constraints. Following the Dirac-Bergman algorithm for constraints, one finds that the canonical brackets are
\bea \label{mag-can-brac1}
\{ \psi_{\alpha(+)} (x), \psi^\dagger_{\beta(-)} (y)\} \Big{|}_{x^+ = y^+} &=&  \frac{i}{\sqrt 2} (\g_+)_{\alpha \beta}\, \delta^\perp (x-y) \,,  \\
\{ \psi_{\alpha(-)} (x), \psi^\dagger_{\beta(+)} (y)\} \Big{|}_{x^+ = y^+} &=&  \frac{i}{\sqrt 2} (\g^+)_{\alpha \beta}\, \delta^\perp (x-y)\,, \label{mag-can-brac2}
\eea
while the diagonal anticommutators vanish
\be
\{ \psi_{\alpha(\pm)} (x), \psi^\dagger_{\beta(\pm)} (y)\} = 0 \,.
\ee
The canonical magnetic Hamiltonian density is given entirely in terms of $\psi_{(+)}$
\be
\mathcal H^{mag} = - \sqrt 2 \, \psi^\dagger_{(+)} (i \g^i\del_i - m)\psi_{(+)}\,.
\ee
With the momentum density defined as 
\be
\mathcal P^{mag}_i =  - \frac{i}{\sqrt 2} \left( \psi^\dagger_{(-)} \gamma^+\del_i \psi_{(+)} +  \psi^\dagger_{(+)} \gamma_+\del_i \psi_{(-)} \right)\,,
\ee
one can easily verify that the commutation relations \eqref{H-P Comm} - \eqref{PP} are satisfied. 
\par
Additionally, the Carroll transformations generated by $G [{\xi^+, \xi^i}] $
\be
\delta \psi_{(+)} = \xi^+ \frac{\del \psi_{(+)}}{\del x^+} + \xi^i \frac{\del \psi_{(+)}}{\del x^i}  - \frac{i}{4} \omega^{ij} \sigma_{ij} \psi_{(+)}\,,
\ee
and $\delta \psi_{(-)} $ as in \eqref{elec-boost} render the magnetic Carroll action \eqref{L-elec} invariant. These two conditions together imply that the magnetic Lagrangian \eqref{L-mag} indeed describes a Carroll-invariant theory in the Hamiltonian sense.
\par
It is important to note that only the $\psi_{(-)}$ modes transform non-trivially under Carroll boosts $c_i$, while the $\psi_{(+)}$ modes remain invariant. This further underscores the fact that the light-cone projection provides a natural and effective distinction between the electric and magnetic modes, both dynamically and through their transformation properties under the light-cone Poincar\'e group and its Carroll subgroup.

%%%%%%%%%%%%%%%%%%%%%%%%%%%%%%%%%%%%%%%%%%%%%%%%%%%%

\subsubsection{Two-point functions}
Varying \eqref{L-mag} with respect to $\psi^\dagger_{(-)}$ and $\psi^\dagger_{(+)}$ respectively yields the equations of motions
\bea
\del_+\psi_{(+)} &=& 0 \,,\\
\del_+ \psi_{(-)} + \g^+ \g^i \del_i \psi_{+} -im \g^+ \psi_{(+)} &=& 0 \,.
\eea
Therefore, in the phase space, $\psi^\dagger_{(-)}$ or equivalently $\pi_{(+)}$ plays the role of a Lagrange mutliplier that enforces the constraint $\dot{\psi}_{(+)} = 0$. The general solutions of these equations are
\bea
\psi_{(+)} (x^+,x^i)&=& \psi^{(0)}_{(+)} (0,x^i)\,, \\
\psi_{(-)} (x^+,x^i) &=& -i\g^+ (i\g^i \del_i +m ) \psi^{(0)}_{(+)} x^+ + \psi^{(0)}_{(-)} (0,x^i)\,, \label{psi-minus}
\eea 
which are characterised by $\psi^{(0)}_{(+)} (0,x^i)$ and $\psi^{(0)}_{(-)} (0,x^i)$.
\par
In order to compute the 2-point function, we recast the magnetic Lagrangian as
\be
\mathcal L^M = \frac{1}{\sqrt 2} \psi^\dagger_M \mathcal D\, \psi_M\,,
\ee
with
\be
\psi_M =\begin{pmatrix}
\psi_{(+)} \\
\psi_{(-)} 
\end{pmatrix} \,,\quad \mathcal D = \begin{pmatrix}
2(i\g^i \del_i - m)& -i\g_+ \del_+\\
-i\g^+ \del_+ & 0
\end{pmatrix} \,.
\ee
From the inverse of the kinetic operator $\mathcal D$, the 2-point functions are found to have the form 
\bea
\label{eqn:gmag}
G_{mag}(\tau,\mathbf x-\mathbf y) =\begin{pmatrix} 
0&i\g^- \\
&\\
i \g^+  \ & -2i \g^-\g^+ \tau  (i\g^i \del_i - m)
\end{pmatrix}\Theta (\tau) \delta^\perp (\mathbf x-\mathbf y) \,,
\eea
where we have used $\tau = x^+-y^+$. Note that the two-point function diverges from the standard power-law behavior expected in the magnetic sector, see e.g~\cite{Chen:2021xkw, Banerjee:2020qjj}. While seemingly unusual, this departure is fully justifiable and can be best understood by first examining the standard canonical analysis.
\par
We now discuss the canonical quantization of this theory. Since $\psi_{(+)}$ is independent of $x^+$, we again consider a mode expansion in the traverse directions alone
\be
   \psi_{(+)}{}_\alpha(x^+, x_\perp) = \int \frac{d^2k}{(2\pi)}\,e^{i \mathbf{k} \cdot \mathbf{x}_\perp}\,  b_\alpha(\mathbf{k}),
   \ee
   and similarly, for $\psi^\dagger_{(+)}$
From \eqref{psi-minus}, we see that the mode expansion for   $\psi_{(-)}{}$contains a homogeneous part $d_\alpha(\mathbf{k})$ and a particular solution depending on $b_\alpha (\mathbf{k})$
\be
    \psi_{(-)}{}_\alpha(x^+,\mathbf{x}_\perp)
    = \int\frac{d^2k}{((2\pi)}\,e^{i\mathbf{k}\cdot\mathbf{x}_\perp}
      \left[
          d_\alpha(\mathbf{k})
          + x^+\,(\g_+)_{\alpha\beta}\,
            (\g^i k_i + m)_{\beta\gamma}\,
            b_\gamma(\mathbf{k})
      \right],
\ee
and similarly, for $\psi^\dagger_{(-)}$. Substituting these into the canonical anticommutators \eqref{mag-can-brac1} - \eqref{mag-can-brac2}, the non-trivial relations among the operators are
\be \label{cross-comm}
\{ b_\alpha(\mathbf{k}), d^\dagger_\beta(\mathbf{k'})\} = \{ d_\alpha(\mathbf{k}), b^\dagger_\beta(\mathbf{k'})\} = i\sqrt 2 (2\pi) (\g_+)_{\alpha \beta} \delta^\perp (\mathbf{k}-\mathbf{k'})\,,
\ee
while all diagonal anticommutators vanish
\be
\{ b_\alpha(\mathbf{k}), b^\dagger_\beta(\mathbf{k'})\} =0 \,, \quad  \{ d_\alpha(\mathbf{k}), d^\dagger_\beta(\mathbf{k'})\} = 0\,.
\ee
The magnetic Hamiltonian in the operator form reads
\be
H^{mag} = \sqrt 2\int\frac{d^2k}{((2\pi)} (\g^i k_i +m )_{\alpha\beta} b_\alpha(\mathbf{k})b^\dagger_\beta(\mathbf{k})\,.
\ee
However, unlike in usual QFT, the $\int b^\dagger b$ cannot serve as a number operator since we have $\{ b, b^\dagger\} = 0$.
\par Some comments on the quantum structure of the magnetic theory are in order. The off-diagonal kinetic structure of $\mathcal{L}^M$ indicates that $\psi_{(+)}$ and $\psi_{(-)}$ form a \textit{symplectic pair} under the Carroll dynamics, rather than two independently quantizable fields. Unlike the electric theory $\mathcal{L}^{\mathrm{elec}}$, where $\psi_{(-)}$ alone is dynamical and the Fock space construction is straightforward, the magnetic theory does not admit a clean decomposition into independent particle and antiparticle sectors in the original basis.
\par
One natural way to address this is to perform a Bogoliubov-like basis rotation, $B = (b+d)/\sqrt{2}$ and $D = (b-d)/\sqrt{2}$, under which the cross-anticommutators~\eqref{cross-comm} give rise to non-vanishing $\{B, B^\dagger\}$ and $\{D, D^\dagger\}$ in the rotated basis. However, diagonalizing the Hamiltonian and constructing a physical Hilbert space with a well-defined vacuum require careful treatment.  This brings us to the issue regarding the specific form of the two-point function (\ref{eqn:gmag}). It is instructive to mention that the departure from the standard power law behaviour is a natural consequence of the quantization scheme. In the literature, the standard Green's functions for the magnetic sector of Carroll CFTs are typically constructed using a highest-weight vacuum---defined by demanding that the state is annihilated by the global symmetry generators---which leads to standard power-law spatial behavior. However, a more natural approach here is to use the notion of an induced vacuum, defined instead by demanding that the canonical lowering operators annihilate the vacuum state. This prescription eliminates the power-law behavior, causing the correlation functions to collapse into a strictly ultra-local form.  Defining the vacuum in this way, however, demands a rigorous construction of the underlying Hilbert space, which likely requires going beyond the standard formulation. 

%%%%%%%%%%%%%%%%%%%%%%%%%%%%%%%%%%%%%%%%%%%%%%%%%%%%

\section{Concluding remarks}

In this work, we construct electric and magnetic actions for Carrollian fermionic via the null reduction of a single Bargmann-invariant Dirac action. Our construction relies on null-reducing the Dirac Lagrangian using light-cone variables and decomposing the spinor into strictly light-cone projections. We utilize this approach for two primary reasons: first, light-cone projections are well-defined in arbitrary spacetime dimensions, whether even or odd. Second, because mass terms explicitly break chiral symmetry, the existing models in the literature are largely restricted to massless cases. Utilizing light-cone projections allows our framework to naturally bypass this limitation and successfully accommodate the massive regime.
\par
By employing light-cone projections, we split the spinor into dynamical `good' and constrained `bad' fermions. We show that the magnetic sector arises naturally from the dynamical modes of the parent Lorentzian theory whereas, the electric sector is obtained by deforming to a Bargmann action which removes the Lorentzian primary constraints and promotes the `bad' fermions to dynamical degrees of freedom. By restricting this deformed higher-dimensional theory onto a null hypersurface and preserving the canonical symplectic structure via a smearing parameter limit $\epsilon \to 0$, we systematically extract the Carroll actions as well as the Carroll Clifford algebra. We validate both the actions through canonical analysis from a Hamiltonian perspective, verifying the Carroll commutation relations, and demonstrating their invariance under Carroll boosts. 
\par
Furthermore, we constructed the two-point functions. The two-point function for the electric sector is in agreement with the well-known form of the Green's function for the electric Carroll sector. However, the magnetic sector involves a fundamental ambiguity: a Hilbert space cannot be defined using standard means. This ambiguity is evident from the fact that the magnetic two-point function exhibits ultra-locality unlike the power law behaviour expected in the magnetic sector. This departure shall not come as a surprise as similar behaviour has been observed for a Carrollian scalar field in the magnetic sector~\cite{Chen:2024voz}. While this contrasts with the power-law behavior traditionally derived using a highest-weight vacuum in Carroll CFTs, it is a direct consequence of adopting a more natural induced vacuum. By defining the vacuum such that it is annihilated by canonical lowering operators, the spatial correlations inherently collapse into an ultra-local form. One possible direction for defining the vacuum is the framework of a rigged Hilbert space, which has been successfully employed in the analogous case of Carrollian scalar field theories~\cite{Chen:2024voz} and may provide the appropriate setting for carrying out the canonical quantization for the present system. Whether this constitutes the full resolution remains to be seen, which we defer to future work. The framework developed in this paper opens several avenues for future investigation. We discuss some of them in the following paragraph
%%%%%%%%%%%%%%%%%%%%%%%%%%%%%%%%%%%%%%%%%%%%%%%%%%%%

\subsection*{Future works}
An immediate extension concerns the quantum field theoretic aspects of these models, particularly the magnetic sector, where the obstruction to defining a standard Hilbert space leads to significant ambiguities. It is worth noting that this difficulty is not unique to fermions: the electric sector of Abelian Carrollian electrodynamics (CED) suffers from the same issue. However, CED remains fundamentally tractable because its tree-level structure is strictly classical, and even upon the introduction of interactions, gauge fixing effectively eliminates loop corrections, preserving this classical nature~\cite{Sharma:2025rug}. As discussed above, for Carroll fermions the situation is more subtle and the Rigged Hilbert space formalism appears to be a natural avenue for addressing this. For interacting theories, the simplest case to consider is a Yukawa-type coupling, which provides a minimal framework to investigate the renormalization properties of these field theories. A rigorous understanding of the quantum structure is also of broader importance, since Carrollian field theories are promising candidates for holographic duals of asymptotically flat gravity, and any such dual must be a genuine quantum field theory. Significant progress has already been made in this direction~\cite{Cotler:2024xhb, Cotler:2025npu, Fredenhagen:2026pia}, and we expect the framework developed here for fermions to serve as a useful step toward that goal. Additionally, it would be interesting to investigate the role of the magnetic sector in a putative holographic dual, given the ultra-locality of its correlation functions. 
\par
Another potential avenue of future research is extending our approach to gauge fields, particularly non abelian sector. Exploring how the light-cone gauge-fixing techniques conventionally employed in QCD map onto the Carrollian limits of gauge and fermionic theories remains an intriguing open question for future study. 
\par
Finally, beyond the inclusion of interactions, the framework developed in this paper invites generalizations to higher-spin fields~\cite{Campoleoni:2021blr,Bekaert:2024itn,DArcy:2026hgl} and spin-3/2 fermions coupled to gravity~\cite{Bergshoeff:2023vfd}, where the dynamics in the parent Lorentzian theory are governed by the Fronsdal action and the Rarita-Schwinger action, respectively. This generalization is particularly compelling for it paves the way to future studies into Carrollian supersymmetry and supergravity~\cite{Bergshoeff:2022iyb, Grumiller:2024dql, Bruce:2026yvw, Bagchi:2026emg}. 

%%%%%%%%%%%%%%%%%%%%%%%%%%%%%%%%%%%%%%%%%%%%%%%%%%%%

 \section*{Acknowledgements}
 We thank Arjun Bagchi, Daniel Grumiller, Saikat Mondal and Romain Ruzziconi for fruitful discussions. SM is grateful for the hospitality of Perimeter Institute where part of this work was carried out. Research at Perimeter Institute is supported in part by the Government of Canada through the Department of Innovation, Science and Economic Development and by the Province of Ontario through the Ministry of Colleges and Universities. This work was supported by a grant from the Simons Foundation (1034867, Dittrich). 
 
 %%%%%%%%%%%%%%%%%%%%%%%%%%%%%%%%%%%%%%%%%%%%%%%%%%%%

% \bibliographystyle{unsrt}
  \bibliographystyle{Biblio}
	\bibliography{Ref}
\end{document}